\documentclass[a4paper,12pt]{article}

\usepackage{latexsym}
\usepackage{graphicx}
\usepackage{rotating}

\usepackage{amsmath,amsfonts}
\usepackage{bm}
\usepackage{epsfig}

\title{Measurements on tones generated in a corrugated flow pipe with special attention to the influence of a low frequency oscillation.}

\author{Ulf R. Kristiansen$^1$, Pierre-Olivier Mattei$^2$, Cedric Pinhède$^2$, Muriel Amielh$^3$ \\
\small $^1$ Acoustics Research Center, NTNU, O.S. Bragstads plass 2b, 7491 Trondheim, Norway\\
\small $^2$ Laboratoire de M\'ecanique et d'Acoustique, CNRS, 31 Chemin Joseph Aiguier, 13009 Marseille, France\\
\small $^3$ Institut de Recherche sur les Ph\'enom\`enes Hors \'Equilibre,  49 rue F. Joliot Curie,  13013 Marseille, France }

\date{In proceedings of the 34th Scandinavian Symposium on Physical Acoustics, Geilo, Norway, 30 January - 2 February 2011. ISBN 978-82-8123-004-0., 2011.}

\begin{document}

\maketitle


\noindent{\bf Summary}

It is well known that an air flow in a corrugated pipe might excite the longitudinal acoustic modes of the pipe. In this letter is reported experiments where a low frequency, oscillating flow with velocity magnitudes of the same order as the air flow   has been added. Depending on the oscillation strength, it might silence the pipe or move the resonances to higher harmonics. It is also shown that a low frequency oscillation by itself might excite a higher frequency acoustic resonance of the pipe.


\section{\label{sec:level1}Introduction}

Sound production in corrugated pipes has been a topic for scientific investigation for almost a century. Short pipes has been used in physics demonstrations and served as musical toys, but also  been  a topic for scientific research. The flow acoustic interaction is complicated as it involves a close interaction between the flow and the acoustics. A review of the literature on the topic up to about 2006 is given by Kristiansen and Wiik \cite{Kristiansen}. The studies up to that time were largely experimental. Simplified models of sound sources interacting with the pipe flow has later been presented by Goyder \cite{Goyder}, Debut et al.\cite{Debut}, and Tonon et al. \cite{Tonon}. In a more direct numerical approach, Popescu and Johansen recently showed, by solving the compressible Navier Stokes equation numerically for a short corrugated pipe, that strong cavity vortices would  interact with the boundary layer in parts of the pipe\cite{Popescu}. No full explanation of the sound generation by the flow  and its  feedback effect has, to the knowledge of the authors,  been published.  

The topic has received renewed interest as the so called "singing riser" problem has become apparent in the natural gas industry. The long flexible pipes used for conveying gas are corrugated on the inside and  are known to exhibit strong sound levels at pure tones on some off shore installations. Some recent publications directly treating the singing riser problem are published by Reinen \cite{Reinen}, Belfroid et al. \cite{Belfroid}. 
In the study by Reinen, a 20m long plastic tube, embedded in concrete, was used. Resonances were established between an open  entry end, and a downstream side branch resonator. With air flowing through the pipe, it would sound at the side branch resonances. In a special test it was found that injecting  sound from an attached loudspeaker, at a frequency one order of magnitude lower, and at a level of the same order as the flow induced resonance, would silence the pipe. 

In the present communication we report the influence of low frequency sound on corrugated pipe resonances in a simple experimental arrangement. A hard walled box has been constructed with two openings, one connected to a small diameter corrugated pipe, and the other to a vacuum cleaner able to draw air through the system and thereby excite the corrugated pipe's longitudinal acoustic resonances. A loudspeaker is attached to one side of the box. For a low frequency pure tone sound from the loudspeaker, the arrangement effectively makes up an excited Helmholtz resonator with the  air in the corrugated pipe oscillating as a single fluid body. For frequencies close to the Helmholtz resonance, high sound pressure levels and  particle velocities (pipe fluid volume oscillations) are expected in the  pipe. The Helmholtz resonance made up by the box and the corrugated pipe is in the present arrangement around 8.8Hz. The vacuum cleaner connecting pipe and different ways of attaching the vacuum cleaner in order to get the right velocity, would slightly change this value. At 10Hz the loudspeaker would give an adequate signal to observe the phenomenon and was therefore chosen as the test frequency. For  a Helmholtz resonator of this type the oscillation velocity is expected nearly constant over the length of the pipe, while the pressure will decrease from the box towards the open end.


\section{Experimental set up}
A sketch of the experimental set up is shown in figure 1. The inner box volume measures 0.29x0.29x0.55 m$^3$. The corrugated pipe has an inner diameter of 25.4{mm, and is 0.64m long. One end is mounted flush with the box's inner surface. The cavity pitch is 5.5 mm. The pipe geometry is similar to the one discussed reference 4.  A 35mm inner diameter pipe connects the box to a vacuum cleaner.  A 240 mm diameter loudspeaker  is positioned at one of the box walls. The loudspeaker was fed 10Hz signals between 0 and 12V. To minimize  static pressure differences on both side of the loudspeaker in a flow situation, it is backed by a 0.07 m$^3$ volume communicating with the box through small holes. The velocity is measured with a Pitot tube at the box side opening of the pipe. The Pitot tube is connected to an electronic manometer of type Alnor AXD. In a special test, a standard hot wire anemometer probe (type 55P11 operated with a DANTEC Streamline CTA system) was used to measure the fluctuating velocity at the pipe's entry section. 
The sound levels were recorded with a 160mm long probe microphone (type G.R.A.S. 40SC), positioned so that it measured the sound levels 30mm into the pipe. A second probe microphone was used as a control microphone in the middle of the tube. The signals were transferred to a computer and analyzed by the dBFA suite 4.9 (01dB Metravib). The microphone was calibrated using standard calibration equipment. The probe corrections are supplied by the manufacturer and were accounted for.

\begin{figure}
\centering 
\includegraphics[width=8cm,height=5cm]{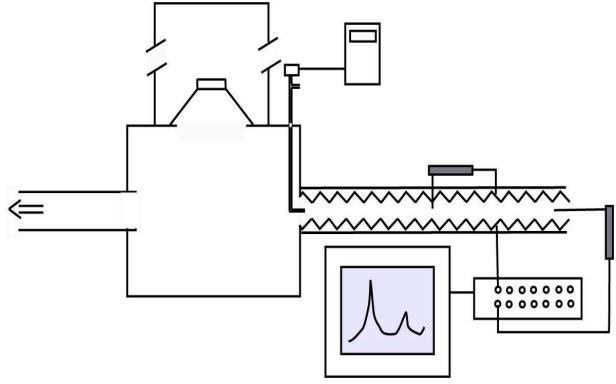}
\caption{A sketch of the experimental set up.}
\label{fig:epsart}
\end{figure}


\section{Results}
\subsection{\label{sec:level2}Flow generated resonances}
In figure 2 are shown the resonances generated by letting the vacuum cleaner draw air through the pipe. The sound pressure level of the dominant peak (measured 30mm into the pipe from the air entry end) is plotted against the velocity measured at the box end of the pipe by the Pitot tube, $U_{pt}$. The fundamental longitudinal resonance could not be excited. By increasing the velocity,  harmonics were found at 506 (square), 751(circle), 1012(star), 1256(diamond), and 1517(plus) Hz.

\begin{figure}
\centering 
\includegraphics[width=8cm,height=6cm]{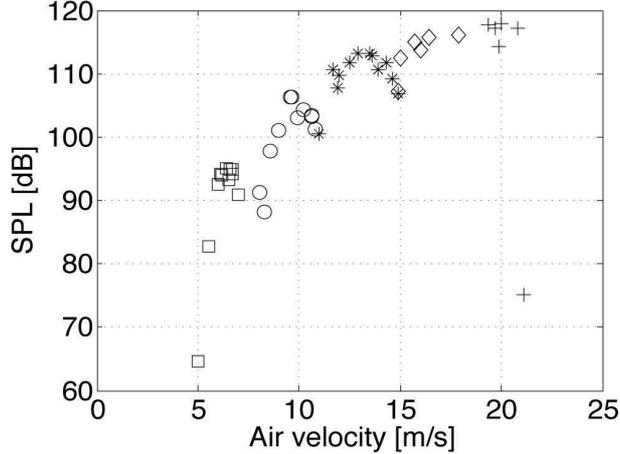}
\caption{Sound pressure levels of dominant frequencies measured at 30mm position as function of flow velocity. }
\end{figure}
\subsection{Resonances generated by 10Hz oscillation alone}

It was also observed that with no air flow drawn through the system, the 10Hz oscillation could by itself excite resonances in the corrugated tube. Figure 3 shows the different dominant peaks as function of $L_{10}$(30), the 10Hz sound pressure level measured 30mm from the flow entry opening. Figure 4 shows the spectrum at a  $L_{10}$(30) level of 107.6dB. We see that the peak is rather broad and modulated by the 10Hz tone.

 In what follows, we chose to represent the oscillating field by a sound pressure level value at a given position, namely the $L_{10}$(30) value.

\begin{figure}[!htb]
\centering 
\includegraphics[width=9cm, height=4cm]{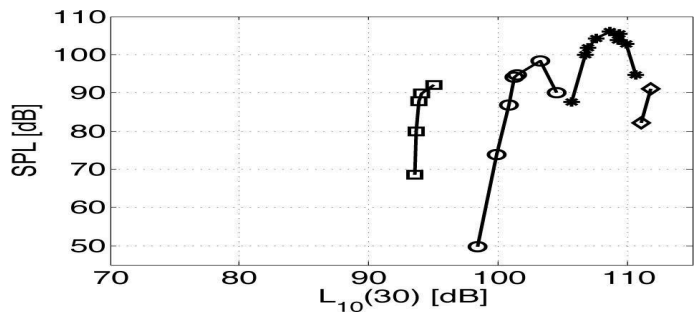}
\caption{ Resonances generated by 10Hz tone alone.}
\end{figure}

\begin{figure}[!htb]
\centering 
\includegraphics[width=9cm, height=4cm]{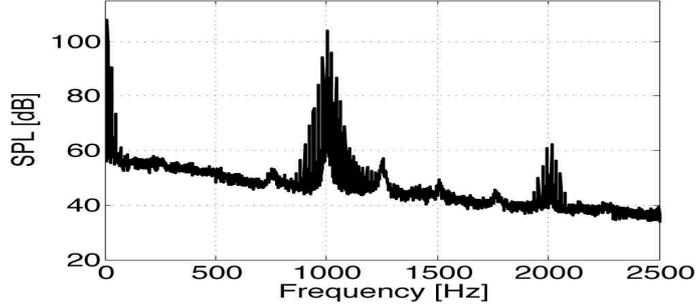}
\caption{ Spectrum measured for $L_{10}(30) =107.6$dB.}
\end{figure}
\subsection{Hot wire measurements}
Some hot wire velocity measurements were done at the pipe's inflow end to record oscillation velocities directly. As no simultaneous velocity and pressure measurements were done, we have not tried to relate velocity values to pressure values at specific positions. The purpose of the velocity measurements were to record order of magnitude values. The  measurements were done in the entry plane of the pipe on the axis position.
Figure 5 shows an example of a fluctuating flow driven by the 10Hz tone alone. The difference in heights  was interpreted as caused by  the difference in exit and entry flows at this position. The exit flow would be more jet like than the entry flow, where air is sucked from all directions. In figure 6 is  plotted the velocity when air is drawn through the system by the vacuum cleaner. The flow is here always entering the pipe which makes the signal more symmetric. The velocity measured with no loudspeaker signal is also plotted. As this signal also contains the flow excited resonance, we can observe the orders of magnitude difference in 10Hz oscillation velocities and the high frequency particle velocities. The difference in the mean level measured by the hot wire and the Pitot value is attributed to the difference in exit and entry flows discussed above.

\begin{figure}[htb]
\centering 
\includegraphics[width=8cm, height=4cm]{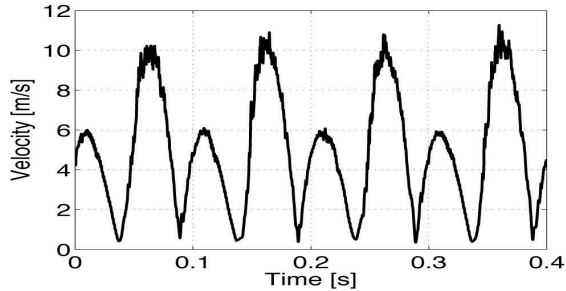}
\caption{ Hot wire measurement at center of the pipe's entry section. $U_{pt}=0\mbox{m/s}$; Vacuum cleaner end closed. Loudspeaker voltage 7.5V, $L_{10}(30)$=101.1dB.}
\end{figure}

\begin{figure}[htb]
\centering 
\includegraphics[width=8cm, height=4cm]{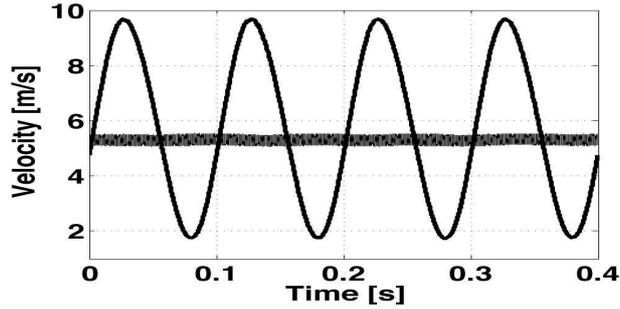}
\caption{ Hot wire measurement at center of entry section, $U_{pt}=10.9\mbox{m/s}$. Loudspeaker voltage 9.0V. $L_{10}(30)=110.5dB$ }
\end{figure}

\subsection{Influence of 10Hz tone on flow generated resonances}
In figure 7 are plotted spectra showing the influence of the 10Hz oscillation on the 2nd. longitudinal pipe resonance. With no 10Hz tone present, this resonance was easily excited with an air flow measured at 6.5m/s with the Pitot tube. Comparing the upper spectra with the one below, it is seen that the resonant peak is reduced at this position by about 40dB with an added $L_{10}$(30) value of 89.4dB. Increasing the tone's level even more, higher order resonances become apparent. It is also seen that these are much broader. A zoom of the lower panel of figure 7 around the 5'th resonance peak shows it to be strongly influenced by the 10Hz signal, see figure 8. The sound levels are measured at a position 30mm into the tube from the air entry end. The sound levels will be  higher at the different resonances' pressure peak positions inside the pipe.

In figures 9,  10, and 11 the 2nd, 3rd and 4th longitudinal resonances were generated by flow velocities $U_{pt }= 6.3, 9.9 \hspace{1mm} \text{and}\hspace{1mm} 12.0 \mbox{m/s}$ respectively. It is seen also here that by adding and increasing the 10Hz signal, the resonance levels are first lowered before the dominant peaks are shifted to the higher harmonics.

\begin{figure}[htb]
\centering 
\includegraphics[width=9cm, height=14cm]{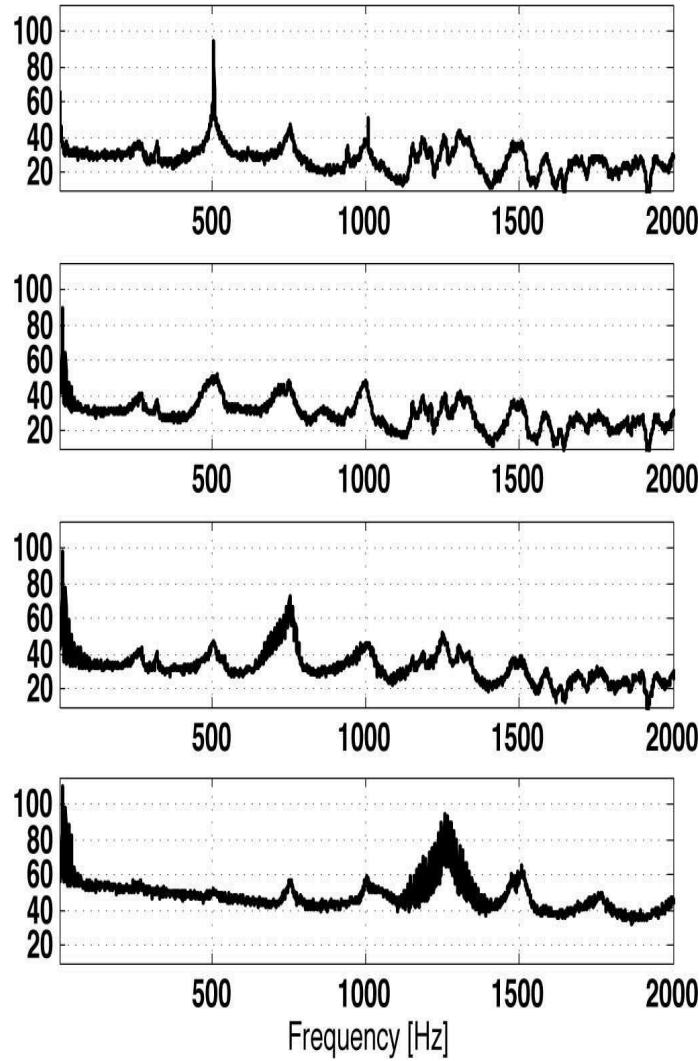}
\caption{ Influence of 10Hz tone on the second longitudinal pipe resonance. Upper figure: resonance generated with flow velocity $U_{pt}=6.5\mbox{m/s}$ (no 10Hz tone). Lower panels show the effect of an added 10Hz tone at increasing levels: ($L_{10}$(30) = 0, 89.4, 96.6, and 109.4dB)}
\end{figure}

\begin{figure}
\centering 
\includegraphics[width=9cm, height=4cm]{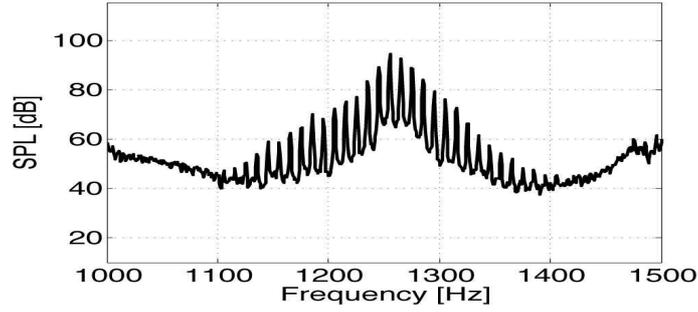}
\caption{ Close up of the resonance peak around 1250 Hz (5th longitudinal resonance) in the lower panel of figure 7.}
\end{figure}

\begin{figure}[h!tb]
\centering 
\includegraphics[width=8.5cm, height=4cm]{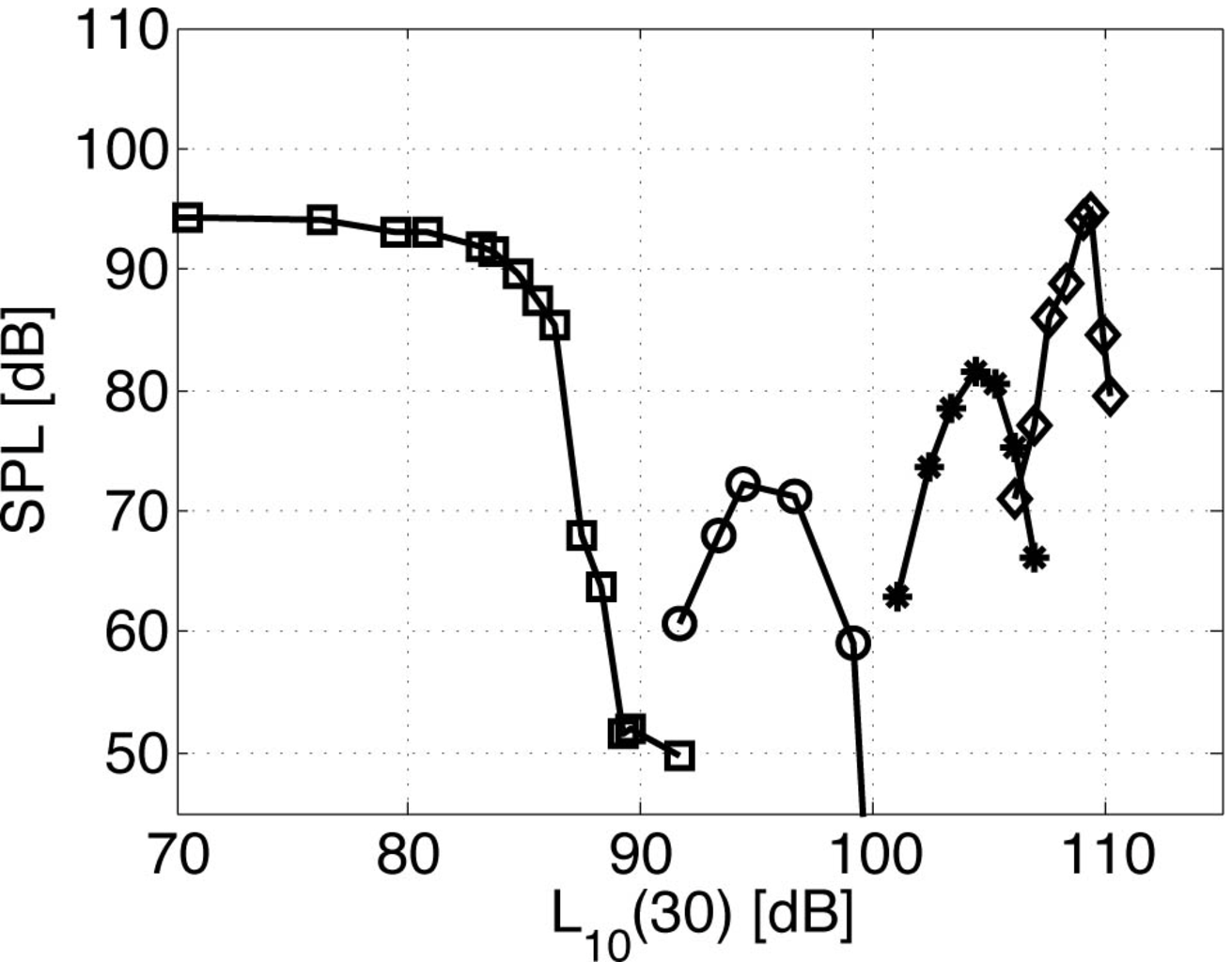}
\caption{ Effect of increasing  10Hz tone level on dominant spectrum peak. Start point is 2nd. harmonic excited with $U_{pt}=6.3\mbox{m/s}$}.
\end{figure}
\begin{figure}[h!tb]
\centering 
\includegraphics[width=8.5cm,height=4cm]{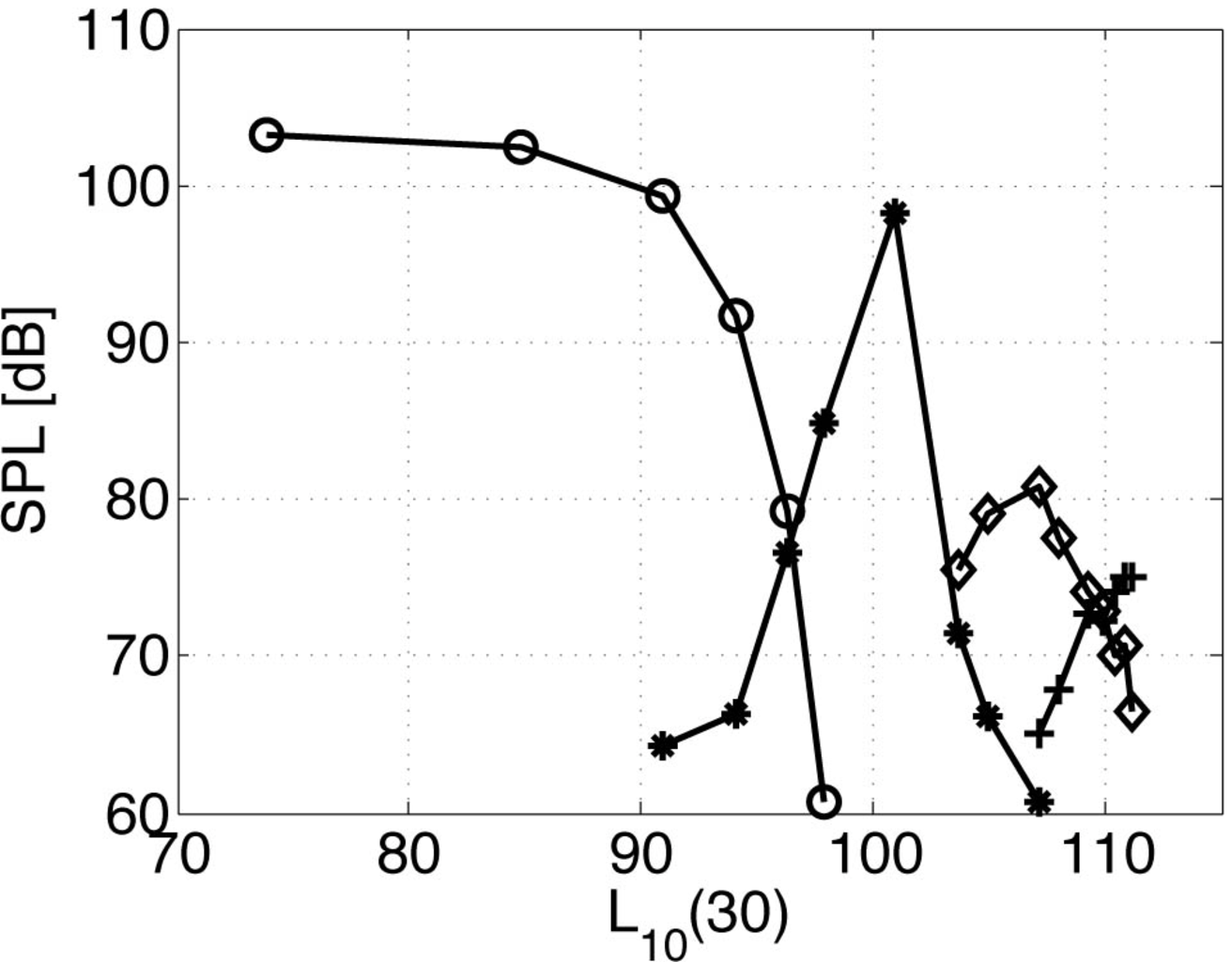}
\caption{Effect of increasing  10Hz tone level on dominant spectrum peak. Start point is 3rd. harmonic excited with $U_{pt}=9.9\mbox{m/s}$.}
\end{figure}
\begin{figure}[h!tb]
\centering 
\includegraphics[width=9cm,height=4cm]{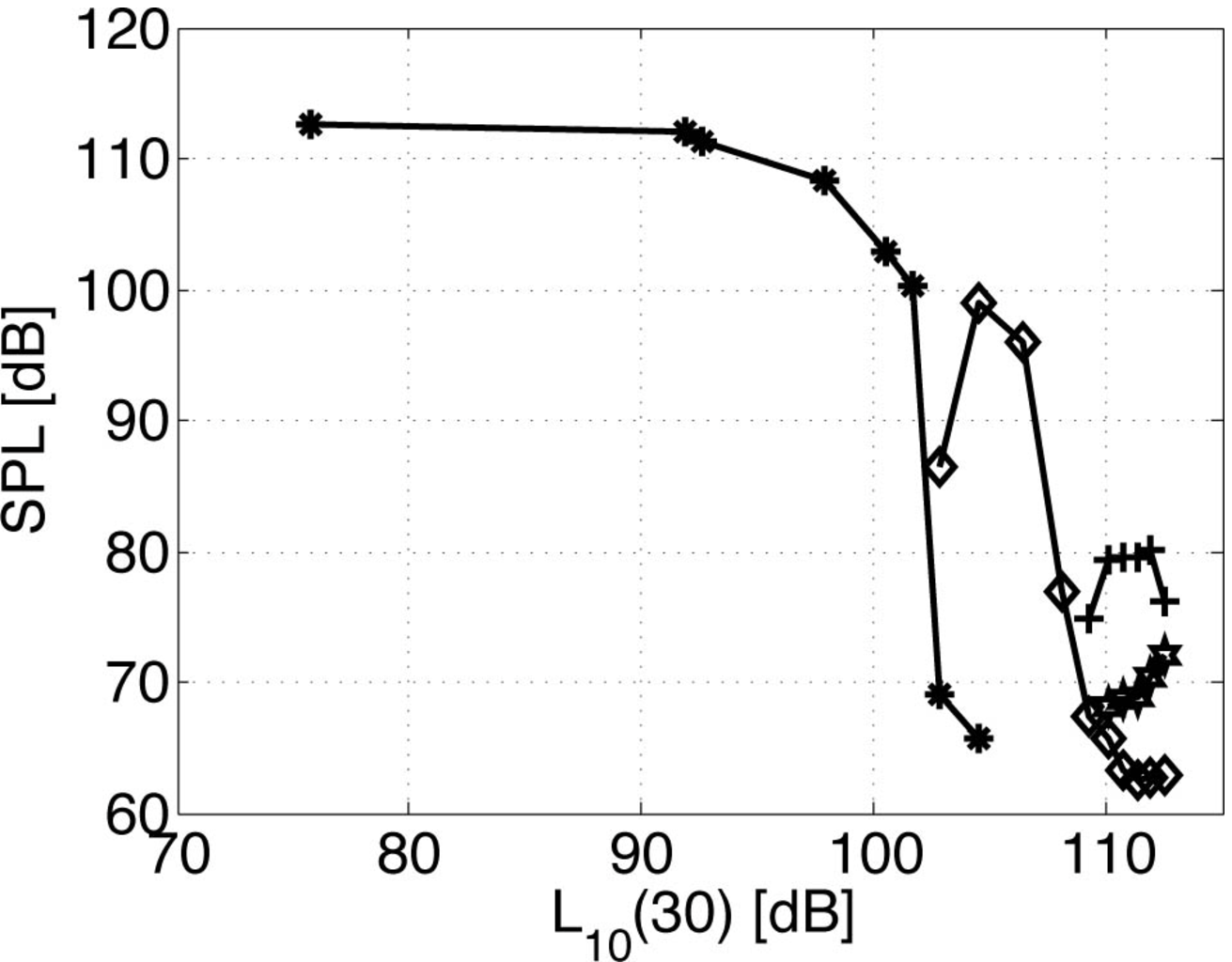}
\caption{ Effect of increasing  10Hz tone level on dominant spectrum peak. Start point is 4th harmonic excited with $U_{pt}=12.0\mbox{m/s}$.}
\end{figure}


\section{Summary}
The experimental results showed that a constant air flow above about 5m/s  would excite the longitudinal acoustic resonances in a 0.64m long, 25mm inner diameter corrugated pipe, and that a superposed oscillating flow, with velocity amplitudes of the same order, would alter the acoustic field in a systematic manner.
The oscillating (10Hz) flow would also by itself excite the acoustic resonances. The oscillating flow was for all measurements represented by the 10Hz sound pressure level measured at a position 30mm from the end of the pipe. For a given steady flow velocity exciting one of the acoustic resonances, the addition of a low frequency flow oscillation  of a certain velocity amplitude would reduce the resonant pressure considerably. A further increase in the oscillation velocity  would then shift the dominant acoustic peak  to the higher resonant frequencies. Some hot wire measurements are presented to show that in the present experiment, the oscillation velocities were of the same order as the constant flow velocities. No detailed measurements have however yet been done on the details of the boundary layer in order to better understand the sound generating mechanism, or  features like for instance the presence of acoustic streaming. We hope to give further information on this in future communications.


\end{document}